\documentclass[entropy,article,accept,pdf,moreauthors]{Definitions/mdpi}
\firstpage{1}
\pubvolume{1}
\issuenum{1}
\articlenumber{0}
\pubyear{2024}
\copyrightyear{2024}
\externaleditor{Sabre Kais and Dominik Szczęśniak}
\datereceived{5 November 2024}
\daterevised{6 December 2024} 
\dateaccepted{17 December 2024}
\datepublished{ }
\hreflink{https://doi.org/} 
\pdfoutput=1 

\Title{W-Class States---Identification and Quantification of Bell-CHSH Inequalities' Violation}

\TitleCitation{W-Class States--- Identification and Quantification of Bell-CHSH Inequalities' Violation}

\Author{Joanna K. Kalaga $^{1,}$*, Wies\l{}aw Leo\'nski $^{1}$ and Jan Pe\v{r}ina Jr. $^{2}$}

\AuthorNames{Joanna K. Kalaga, Wies\l{}aw Leo\'nski and Jan Pe\v{r}ina Jr.}

\AuthorCitation{{Kalaga}, J.K.; Leo\'nski, W.; Pe\v{r}ina Jr., J.}

\address{%
$^{1}$ \quad Quantum Optics and Engineering Division, Institute of Physics, University of Zielona G\'ora, Prof.~Z.~Szafrana 4a, 65-516 Zielona G\'ora, Poland; w.leonski@if.uz.zgora.pl
\\
$^{2}$ \quad {Joint}  {Laboratory}  of Optics of Palack\' y {University} and {Institute} of Physics of AS CR, {Faculty} of Science, Palack\'{y} {University}, 17. listopadu 12, 779~00 Olomouc, Czech Republic; jan.perina.jr@upol.cz
}
\corres{Correspondence: j.kalaga@if.uz.zgora.pl
}

\abstract{We discuss a family of W-class states describing three-qubit systems. For such systems, we analyze the relations between the entanglement measures and the nonlocality parameter for a two-mode mixed state related to the two-qubit subsystem. We find the conditions determining the boundary values of the negativity, parameterized by concurrence, for violating the Bell-CHSH inequality. Additionally, we derive the value ranges of the mixedness measure, parameterized by concurrence and negativity for the qubit--qubit mixed state, guaranteeing the violation and non-violation of the Bell-CHSH inequality.}


\begin{document}

\section{Introduction}
Nonlocality, along with entanglement and quantum steering, is a phenomenon that we do not observe in classical systems. The nonlocality of quantum mechanics was described by A. Einstein, B. Podolski, and N. Rosen in their paper ``Can quantum-mechanical description of physical reality be considered complete?''~\cite{EPR35}. However, discussions of nonlocality predate this famous work on the EPR paradox. Physicists have been concerned with the concept of nonlocality since the formulation of the Copenhagen interpretation of quantum mechanics by Werner Heisenberg, Niels Bohr, and others in the 1920s. In 1927, during the 5th Solvay Conference, A. Einstein emphasized the nonlocality associated with the reduction of the wave function of a single particle and drew attention to its implications in relation to the Copenhagen interpretation of quantum mechanics~\cite{Solvay27}.

In 1964, J.~S.~Bell~\cite{B64} introduced an inequality known as Bell’s inequality, which allows us to identify the nonlocality in two-partite physical systems. Following this, research on the nonlocality of quantum systems continued to develop. In 1969, J. F. Clauser, M. A. Horne, A. Shimony, and R. A. Holt generalized the Bell theorem by introducing an inequality that is now called the Bell-CHSH inequality~\cite{CHS69}. The generalization of Bell's theory by Clauser, Horne, Shimony, and Holt triggered the development of research on nonlocality and quantum correlations.

One of the first experiments testing the Bell inequality was carried out by J. Clauser \cite{FC72} and A. Aspect \cite{AGR82}. In the following years, the experimental research on the Bell inequality was conducted by, among others, A. Zeilinger \cite{WJS98, GVW15}, D. J. Wineland \cite{RKM01}, R. Hanson \cite{HBD15}, and L. Shalm \cite{SMC15}. These experiments confirmed that the Bell inequality is violated and quantum mechanics do not satisfy the principles of local realism.

Quantum states that violate the Bell inequality have found a wide range of applications, and nonlocality, as the form of the strongest quantum correlations, is suitable as a resource for quantum information processing. The Bell-nonlocal states are strongly correlated, and the Bell nonlocality is stronger than the entanglement or the quantum steering~\cite{WJD07,JWD07}.
Therefore, states that violate the Bell inequality are utilized in various fields, including quantum communication and cryptography~\cite{H75b,E91,AGM06,ABG07,AMP06,BLC17,ZKC19,ZFK20}. Despite several decades having passed since the foundational papers by A. Einstein, B. Podolski, and N. Rosen, as well as J. Bell, nonlocality remains a highly attractive topic for research in modern physics~\cite{LX05,DC18,TWF18,FGH19,FDK20,BC20,SGB21,KMF21,VZS22}.

In this paper, we concentrate only on the states categorized as W-class states. Due to their broad applicability, W states have received considerable attention in recent research~\cite{KS08,AG09,CCL11,LZZ20,SC20}. These states are widely used in quantum information systems, such as cryptographic protocols \cite{JQC07,CWG02}, systems for quantum teleportation \cite{ST02a,JP02,ST02b}, or dense coding~\cite{YBM17,ZWL18,RCD18}. Therefore, our analysis of W-class states is particularly important because of their potential applications.

This paper is organized as follows: In Section~\ref{sec_system}, we describe two families of states of a three-qubit system that are categorized as W-class states. In Section~\ref{sec_concurrence}, we derive the formulas giving the parameters characterizing nonlocality for the system with one and two excitations. Then, we discuss the possibility of generation two-qubit states violating the Bell-CSHS inequality and analyze the relations between the nonlocality parameters. Applying suitable entanglement measures, we identify the ranges of the values of negativity, parameterized by concurrence, for qubit--qubit states that assure the violation or fulfillment of the Bell-CSHS inequality. In Section~\ref{sec_entropy}, we study the relationship between the nonlocality parameter and the linear entropy that quantifies the mixedness of the states. We find conditions for the degree of mixedness that imply the violation of the Bell-CSHS inequality. Finally, in Section~\ref{sec_summary}, we present our conclusions.

\section{The Three-Qubit System}
\label{sec_system}
In this study, we pay attention to the nonlocality properties of the states that describe three-qubit systems. In particular, we focus on the mutual relation between the nonlocality parameter and the entanglement measures, such as negativity and concurrence. In our investigation, we consider two cases.

In the first case, the total number of photons/phonons in a system without interaction with the environment is $\langle\hat{n}\rangle=\langle\hat{n}_1\rangle+\langle\hat{n}_2\rangle+\langle\hat{n}_3\rangle=1$, and the qubits considered here are labeled $1$, $2$, and $3$. For such a system, we can write the wave function of the state~as

\begin{equation}
 \vert \psi\rangle= C_{001}\vert 001\rangle+C_{010}\vert 010\rangle+C_{100}\vert 100\rangle.
\label{psiI}
\end{equation}

The corresponding density matrix is derived as
\begin{linenomath}
\begin{equation}
\rho = \vert \psi\rangle \langle \psi\vert=
\left[ \begin{array}{cccccccccc}
 0 & 0 & 0 & 0 & 0 & 0 & 0 & 0 \\
 0 & P_{001} & C^{*}_{001} C_{010} & 0 & C^{*}_{001} C_{100} & 0 & 0 & 0 \\
 0 & C^{*}_{010} C_{001} & P_{010} & 0 & C^{*}_{010} C_{100} & 0 & 0 & 0 \\
 0 & 0 & 0 & 0 & 0 & 0 & 0 & 0 \\
 0 & C^{*}_{100} C_{001} & C^{*}_{100} C_{010} & 0 & P_{100} & 0 & 0 & 0 \\
 0 & 0 & 0 & 0 & 0 & 0 & 0 & 0 \\
 0 & 0 & 0 & 0 & 0 & 0 & 0 & 0 \\
 0 & 0 & 0 & 0 & 0 & 0 & 0 & 0
\end{array} \right] ,
\label{rhoI}
\end{equation}
\end{linenomath}
where the complex probability amplitudes $C_{ijk}$ correspond to the states $\vert ijk\rangle$, and $P_{ijk}=C_{ijk}^{*} C_{ijk}$ are the probabilities.

Similarly, in the second analyzed case, $\langle\hat{n}\rangle$ equals $2$. For such a situation, the wave function describing the system state takes the following form:
\begin{linenomath}
\begin{equation}
\vert \psi\rangle= C_{011}\vert 011\rangle+C_{101}\vert 101\rangle+C_{110}\vert 110\rangle .
\label{psiII}
\end{equation}
\end{linenomath}

Below, we call the system in the state described by the wave function~(\ref{psiII}) 'the doubly excited system' or 'the system with double excitation'. Contrary to this, the system with the wave function~(\ref{psiI}) [$\langle\hat{n}\rangle=1$] is referred to as 'the single excited system' or 'the system with single~excitation'.

The states described by Equations~(\ref{psiI}) and (\ref{psiII}) belong to the same class of states \cite{AAJ01,SG08,EDZ18}, and we obtain the same results for the single and double excitations in the system. For this reason, in the rest of the paper, we only present the calculation for the single excitation in the system. \textls[-5]{The appropriate formulas defining the bipartite concurrence, negativity, linear entropy, and nonlocality parameter are found in turn in \mbox{Equations (\ref{nonlocality_paramIa}), (\ref{nonlocality_paramIb}), (\ref{C_ij_probab}), (\ref{N_ij_probab}), and~(\ref{eq_E})}}. To obtain the formulas that are valid for the doubly excited system, we have to exchange $0 \iff 1$.

\section{The Concurrence, Negativity, and Degree of Nonlocality}
\label{sec_concurrence}

The analysis presented here is an extension of our previous work~\cite{KL17} and, in some sense, the work by Wen-Yang Sun et al. \cite{SWF18}. Although Wen-Yang Sun et al. analyzed the Bell-type nonlocality for W-type states, they focused only on the relations between the nonlocality parameter and the degree of coherence. In contrast to that paper \cite{SWF18}, we study here the relation among the nonlocality parameters for three pairs of qubits (1-2, 2-3, 1-3) and precisely define the relations among them (like-monogamy relation). Additionally, we establish the relation between the entanglement measures (concurrence and negativity) for the Bell-type nonlocal states. We also find the boundary values for both concurrence and negativity, above which all states violate the Bell-CSHS inequality. We derive the relations between linear entropy and entanglement measures for the Bell-type nonlocal states. We focus on W-states because they have a wide range of applications in quantum teleportation systems, dense coding, and cryptographic protocols.

In this paper, we study the relation between the entanglement and the Bell nonlocality for the two families of states of a three-qubit system that are categorized as W-class states. These states do not exhibit genuine tripartite entanglement. In 2000, Coffman et al. \cite{CKW00} proved that, for the states analyzed by us, the concurrences satisfy the relation $C^2_{ij}+C^2_{ik}=C^2_{i(jk)}$, where $C_{ij}$ and $C_{ik}$ quantify the entanglement between two qubits, and $C_{i(jk)}$ is a measure of the entanglement between qubit $i$ and two others. So, we see that $C^2_{i(jk)}$ does not extend beyond two-qubit entanglements. Therefore, in our further analysis, we only focus on correlations occurring in two-qubit subsystems. Primarily, we reveal mutual relations between various quantities describing correlations in the quantum system. Nonlocality is quantified by the Bell parameter $B_{CHSH}$ which is introduced in the Bell-CHSH inequality \cite{CHS69}:
\begin{linenomath}
\begin{equation}
\vert Tr \left(\rho B_{CHSH} \right) \vert \leq 2 .
\label{CHSHineq}
\end{equation}
\end{linenomath}

It takes the following form:
\begin{linenomath}
\begin{equation}
B_{CHSH}= \mathbf{a} \cdot \sigma \otimes \left( \mathbf{b}+\mathbf{b'} \right) \cdot \sigma + \mathbf{a'} \cdot \sigma \otimes \left( \mathbf{b} -\mathbf{b'} \right) \cdot \sigma .
\end{equation}
\end{linenomath}
where $\mathbf{a}$, $\mathbf{a'}$, $\mathbf{b}$, and $\mathbf{b'}$ are the unit vectors in $\Re^3$, $\sigma $ is the vector of Pauli spin matrices $\sigma_i$ $(i=1,2,3)$, and $\mathbf{a} \cdot \sigma$ is the scalar product defined as $\sum_{i=1}^3 a_i\sigma_i$.

It is known that the density matrix $\rho$ of any state can be expressed, on the Hilbert--Schmidt basis, as
\begin{linenomath}
\begin{equation}
\rho =\frac{1}{4}\left( I\otimes I +\mathbf{r} \cdot \sigma\otimes I + I \otimes \mathbf{s}\cdot \sigma +\sum_{n,m=1}^{3}t_{n,m}\sigma_n \otimes \sigma_m \right) ,
\label{rho}
\end{equation}
\end{linenomath}
where $\mathbf{r}$ and $\mathbf{s}$ are the unit vectors in $\Re^3$, and the coefficients $t_{n,m}=Tr \left(\rho\sigma_n \otimes \sigma_m \right)$ form a matrix $T_p$.
The measure of violation of the Bell-CHSH inequality (\ref{CHSHineq}) was introduced by Horodecki~et al. \cite{HHH95,H96}:
\begin{linenomath}
\begin{equation}
M(\rho)= \max_{k<l}\{u_k+u_l\} ,
\end{equation}
\end{linenomath}
where $u_k$ $(k=1,2,3)$ are the eigenvalues of the real symmetric matrix $U_{\rho}=T^T_pT_p$ ($T^T_p$ is the transposition of matrix $T_p$). The Bell-CHSH inequality (\ref{CHSHineq}) is violated by the matrix $\rho$ given in Equation~(\ref{rho}) if the nonlocality parameter $M(\rho)>1$. For the system described by the wave function $\vert \psi\rangle$ given in Equation~(\ref{psiI}) with one excitation present, we express the nonlocality parameters for each pair of qubits using the probabilities $P$. First, we find a reduced density matrix $\rho_{ij}$ representing the state of the two-qubit subsystems. This matrix is derived from the full three-qubit density matrix by tracing out one subsystem $\rho_{ij}=Tr_{k}(\rho_{ijk})$ $-$ the qubit $k$. The indices $i$, $j$, and $k$ label the qubits $(i,j,k=1,2,3)$. For instance, when we trace out subsystem $3$, we find $\rho_{12}$:
\begin{linenomath}
\begin{equation}
{\rho_{12}=Tr_3(\rho_{123})} =
\left[ \begin{array}{cccc}
 P_{001} & 0 & 0 & 0 \\
 0 & P_{010} & C^{*}_{010} C_{100} & 0  \\
 0 & C^{*}_{100} C_{010} & P_{100} & 0 \\
 0 & 0 & 0 & 0
\end{array} \right] .
\label{ro12}
\end{equation}
\end{linenomath}

Next, the matrix $T_p$ is derived as
\begin{linenomath}
\begin{equation}
T_p=
\left[ \begin{array}{ccc}
 C^{*}_{010} C_{100}+C^{*}_{100} C_{010} & -i C^{*}_{010} C_{100}+iC^{*}_{100} C_{010} & 0 \\
 i C^{*}_{010} C_{100}-iC^{*}_{100} C_{010} & C^{*}_{010} C_{100}+C^{*}_{100} C_{010} & 0  \\
 0 & 0 &  \left(P_{001}-P_{010}-P_{100} \right)^2
\end{array} \right] ,
\label{Tp}
\end{equation}
\end{linenomath}
and the matrix $U_{\rho_{12}}$ is found as follows:
\begin{linenomath}
\begin{equation}
U_{\rho_{12}}=T^T_pT_p=
\left[ \begin{array}{ccc}
 4P_{010}P_{100} & 0 & 0 \\
 0 & 4P_{010}P_{100} & 0  \\
 0 & 0&  \left(P_{001}-P_{010}-P_{100} \right)^2
\end{array} \right] .
\label{U}
\end{equation}
\end{linenomath}

According to Equation~(\ref{U}), the three eigenvalues of matrix $U_{\rho_{12}}$ are obtained as follows:
\begin{linenomath}
\begin{eqnarray}
u_1 &=& u_2 = 4P_{010}P_{100} ,\nonumber \\
u_3 &=& \left(P_{001}-P_{010}-P_{100} \right)^2.
\end{eqnarray}
\end{linenomath}

The nonlocality parameter then takes the following form:
\begin{linenomath}
\begin{eqnarray}
\label{nonlocality_paramIa}
 M(\rho_{12})&=& \max \left\{ \left(P_{001}-P_{010}-P_{100} \right)^2 + 4P_{010}P_{100}, \; 8P_{010}P_{100}\right\} .
\end{eqnarray}
\end{linenomath}

Considering other pairs of qubits, we derive the following formulas for the nonlocality~parameter:
\begin{linenomath}
\begin{eqnarray}
\label{nonlocality_paramIb}
M(\rho_{13})= \max \left\{ \left(P_{010}-P_{001}-P_{100} \right)^2 + 4P_{001}P_{100}, \; 8P_{001}P_{100} \right\} ,\nonumber \\
M(\rho_{23})= \max \left\{ \left(P_{100}-P_{010}-P_{001}
\right)^2 + 4P_{010}P_{001}, \; 8P_{010}P_{001}\right\} .
\end{eqnarray}
\end{linenomath}

\begin{figure}[H]
\centering
\hspace{-15em}\includegraphics[width=0.6\textwidth]{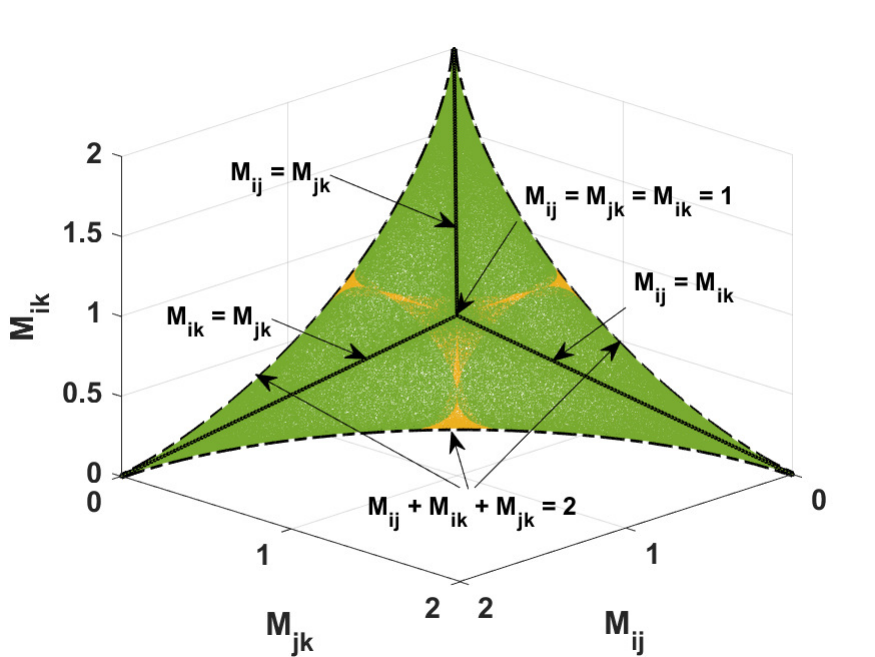}
\caption{Relation between the nonlocality parameters for the states violating (green area) and not violating (yellow area) the Bell-CSHS inequality established numerically. Black curves (solid and dot-dashed) are plotted according to the derived formulas.}
\label{fig_1}
\end{figure}

\begin{figure}[H]
\centering
\includegraphics[width=0.495\textwidth]{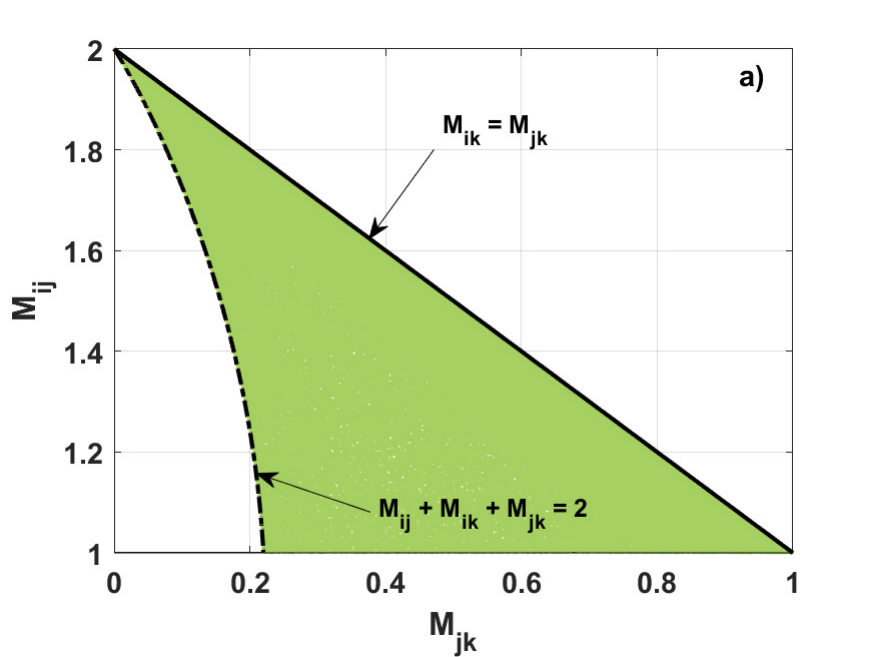}
\includegraphics[width=0.495\textwidth]{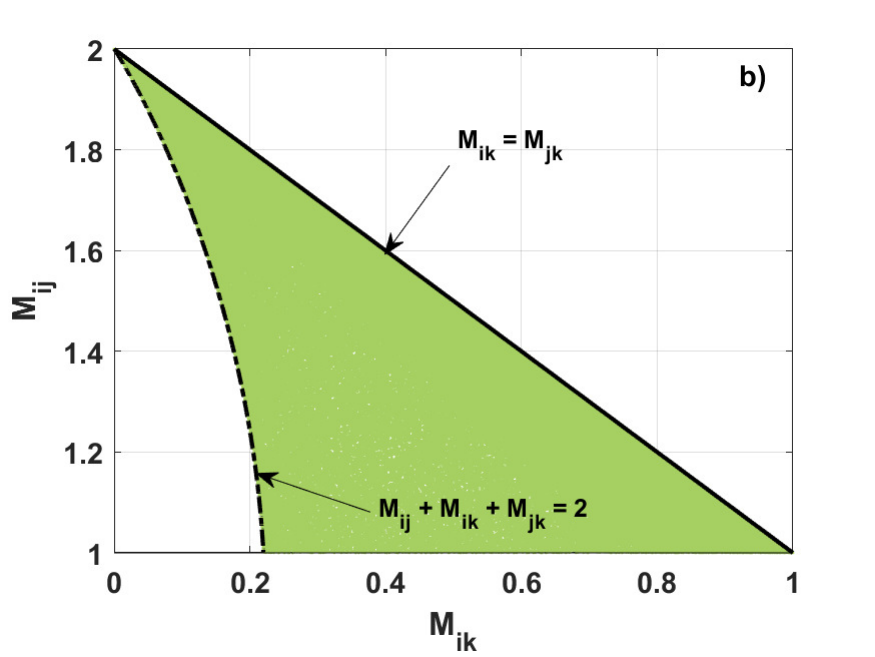}
\caption{{Relation}  between the nonlocality parameters for the states violating the Bell-CSHS inequality established numerically: (\textbf{a}) $M_{ij}$ vs. $M_{jk}$, (\textbf{b}) $M_{ij}$ vs. $M_{ik}$. Black curves (solid and dot-dashed) are plotted according to the derived formulas. }
\label{fig_2}
\end{figure}

The diagrams depicting the mutual relations between the nonlocality parameters for the system with one excitation are plotted in Figures~\ref{fig_1} and \ref{fig_2}. In Figure~\ref{fig_1}, the green region corresponds to the three-qubit states for which at least one of the three nonlocality parameters is greater than unity. The yellow area is related to the states where all nonlocality parameters are smaller or equal to unity. Note that Figure \ref{fig_2}a,b are similar because of the symmetry in Figure~\ref{fig_1}.

The presented results were obtained numerically. We randomly generated $\sim10^6$ three-qubit states defined by the density matrix $\rho$ given in Equation~(\ref{rhoI}). Next, we found the two-qubit density matrices by tracing out one of the three subsystems, and, for each qubit--qubit state, we calculated the nonlocality parameters $M(\rho_{ij})$. In our further considerations, we simplify the notation by writing $M_{ij}$ instead of $M(\rho_{ij})$.

Following Figures~\ref{fig_1} and \ref{fig_2}, we see that, once one of the three nonlocality parameters is greater than one, the other two parameters attain values smaller or equal to unity. Thus, if one two-qubit state violates the Bell-CHSH inequality, then the other two-qubit states cannot violate this inequality. Additionally, $M_{ij}$ reaches the maximal value for a given $M_{jk}$ if $M_{ik}=M_{jk}$ (see the black solid line in Figures~\ref{fig_1} and \ref{fig_2}). For such a case, the relation between the nonlocality parameters takes the following form:
\begin{linenomath}
\begin{equation}
M_{ij}=2-M_{jk} \quad for \quad M_{jk} = M_{ik} .
\label{eq_M1}
\end{equation}
\end{linenomath}

Interestingly, if one of the three nonlocality parameters takes the maximal possible value ($M_{ij}=2$), the other two parameters are equal to zero. Moreover, the relation (14) is satisfied when the two-qubit state described by the reduced density matrix $\rho_{ij}$ is pure. This situation arises when one of the probabilities ($P_{001}$, $P_{010}$, or $P_{100}$) is equal to zero, and the remaining probabilities are related to parameter $M_{jk}$ in a specific manner, which can be described as follows:
\begin{linenomath}
\begin{equation}
{\rm If} \qquad P_{001}=0 \qquad {\rm then}\qquad
\{P_{010};P_{100}\}=\left\{\frac{1}{2}\left(1\pm\sqrt{M_{jk}}\right);\frac{1}{2}\left(1\mp\sqrt{M_{jk}}\right)\right\}.
\end{equation}
\end{linenomath}

Now, we address the boundary for states that violate the Bell-CHSH inequality represented by the dash-dotted line drawn in Figure~\ref{fig_2}a. For $M_{jk} < \left(1-\sqrt{8\sqrt{2}-11} \right)/2$, the minimal achievable degree of nonlocality parameter $M_{ij}$ is given by the following formula:

\begin{linenomath}
\begin{eqnarray}
M_{ij}&=&\frac{1}{16}\left(2-\sqrt{4+4\sqrt{2M_{jk}}-6M_{jk}} +\sqrt{2M_{jk}}\right)\left\{\sqrt{4+4\sqrt{2M_{jk}}-6M_{jk}}\right.\nonumber \\
&&+ 6-\sqrt{2M_{jk}} -2\left[2-\sqrt{4+4\sqrt{2M_{jk}}-6M_{jk}} \right.\\
&&\left.\left. +\left(3\sqrt{2+2\sqrt{2M_{jk}}-3M_{jk}}-2\sqrt{2}\right)\sqrt{M_{jk}}+3M_{jk}\right]^{1/2}\right\}. \nonumber
\label{eq_M2}
\end{eqnarray}
\end{linenomath}

The same relation is derived for the parameters $M_{ij}$ and $M_{ik}$ plotted in Figure~\ref{fig_2}b. Additionally, the states corresponding to the dash-dotted line in Figures~\ref{fig_1} and \ref{fig_2} are the states that fulfill the relation $M_{ij}+M_{jk}+M_{ik}=2$.

In our further analysis, we concentrate on finding the relation between the entanglement and the Bell nonlocality parameter for two distinct groups of states that (a) violate and (b) fulfill the Bell-CHSH inequality with one and two excitations. To quantify the entanglement in a bipartite system, we use the concurrence $C_{ij}$ \cite{HW98, W98}:
\begin{linenomath}
\begin{equation}
C_{ij}=C(\rho_{ij})=\max\left(\sqrt{\lambda_{I}}-\sqrt{\lambda_{II}}-\sqrt{\lambda_{III}}-\sqrt{\lambda_{IV}},0
\right)\, .
\label{concurrence}
\end{equation}
\end{linenomath}

The eigenvalues $\lambda_{l}$ characterize the matrix $R$ obtained along the relation $R=\rho_{ij}\tilde{\rho}_{ij}$, where $\tilde{\rho}_{ij}$ is defined as $\tilde{\rho}_{ij}=\sigma_{y}\otimes\sigma_{y}\rho_{ij}^{*}\sigma_{y}\otimes\sigma_{y}$, and $\sigma_{y}$ is the $2\times 2$ Pauli matrix.

The negativity $ N_{ij} $, the second applied measure of the entanglement, is defined by the following formula~\cite{P96,HHH96}:
\begin{linenomath}
\begin{equation}
N_{ij}=N(\rho_{ij})=\max\left(0,\, -2 \min_{l} \lambda_{l}\right) ,
\label{negativity}
\end{equation}
\end{linenomath}
where $\lambda_{l}$ are the eigenvalues after partial transposition of the matrix $\rho_{ij}$.

\begin{figure}[H]
\centering
\hspace{-15em}\includegraphics[width=0.6\textwidth]{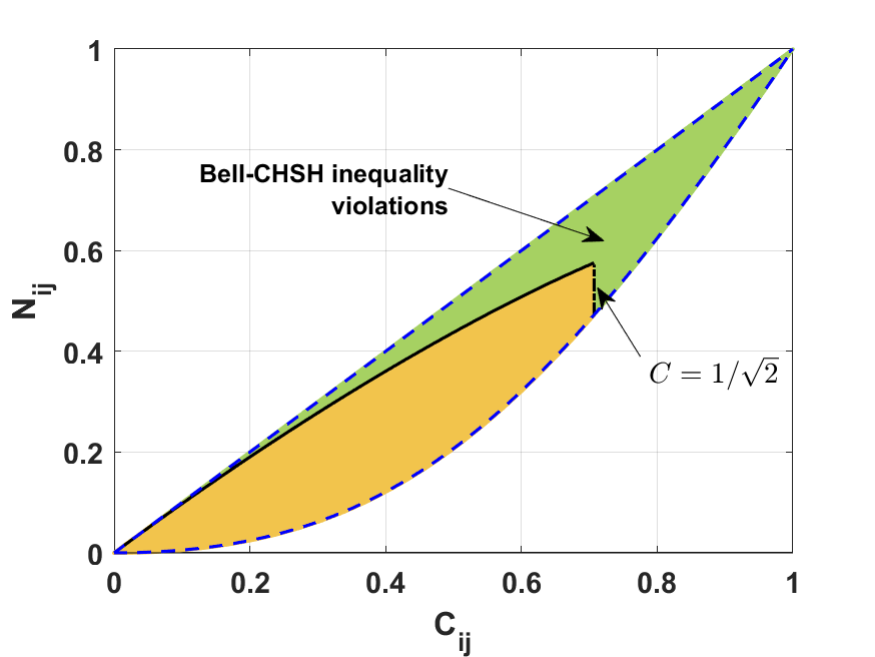}
\caption{Negativity $N_{ij}$ versus concurrence $C_{ij}$ for the states violating (green area) and fulfilling (yellow area) the Bell-CSHS inequality established numerically. Black curves (solid and dot-dashed) are plotted according to the derived formulas. Blue dashed curves are plotted applying the results given in \cite{VAD01}.}
\label{fig_3}
\end{figure}

Figure~\ref{fig_3} reveals a mutual relation between the two measures of entanglement, i.e., the negativity and the concurrence. The green region corresponds to the states that violate the Bell-CHSH inequality, whereas the yellow area represents the states that fulfill the inequality. The black solid and dot-dashed lines represent the border between these two groups of states. The blue dashed lines are related to the upper and lower bounds of the negativity. The minimal and maximal possible values of the negativity corresponding to a given concurrence were found in 2001 by Verstraete et al.~\cite{VAD01}. They showed that, for two-qubit mixed states, the negativity $N_{ij}$ takes values smaller or equal to the given concurrence $C_{ij}$. Moreover, the minimal values of the negativity are equal to $\sqrt{(1-C_{ij})^2 + C_{ij}^2}-(1-C_{ij})$, and the negativity attains this value for the states whose degree of entanglement cannot be increased by any global unitary operation. Such states, known as the Werner states \cite{W89}, are the two-qubit mixed states described as mixtures of Bell and separable states~\cite{IH00}.

To find the upper bound of the negativity for a state not violating the Bell-CSHS inequality, we express the concurrence and the negativity in terms of the probabilities $P_{ijk}$. Applying the formula in Equation~(\ref{concurrence}), we obtain the expressions for the concurrence for different pairs of qubits (more details are found in \cite{KL17}):
\begin{linenomath}
\begin{eqnarray}
C_{12}=\sqrt{4P_{100}P_{010}} ,\nonumber \\
C_{13}= \sqrt{4P_{100}P_{001}} , \nonumber \\
C_{23}=\sqrt{4P_{010}P_{001}} .
\label{C_ij_probab}
\end{eqnarray}
\end{linenomath}

Using Equation~(\ref{negativity}), we find the negativity \cite{KL17}:
\begin{linenomath}
\begin{eqnarray}
N_{12}=\sqrt{P_{001}^2+4P_{100}P_{010}}-P_{001} , \nonumber \\
N_{13}=\sqrt{P_{010}^2+4P_{100}P_{001}}-P_{010} ,\nonumber \\
N_{23}=\sqrt{P_{100}^2+4P_{010}P_{001}}-P_{100} .
\label{N_ij_probab}
\end{eqnarray}
\end{linenomath}

In the next step, we compare the concurrence, negativity, and Bell nonlocality parameter. We know that, for the states that fulfill the Bell-CSHS inequality, the maximal value of the Bell nonlocality parameter is equal to unity. Knowing that the sum of all probabilities is normalized, and applying Equations~(\ref{nonlocality_paramIa}), (\ref{nonlocality_paramIb}), (\ref{C_ij_probab}), and (\ref{N_ij_probab}), we can obtain the formula determining the maximal possible value of negativity for a given value of the concurrence for the states violating the Bell-CSHS inequality. Thus, for the states with the nonlocality parameter $M_{ij}>1$, the negativity satisfies the following inequalities that involve the concurrence:
\begin{linenomath}
\begin{eqnarray}
\frac{1}{2}\left( -1 +\sqrt{1-C_{ij}^2}+\sqrt{2+3C_{ij}^2-2\sqrt{1-C_{ij}^2}} \right)< N_{ij} \leq C_{ij} \quad {\rm for} \quad C_{ij}\leq \frac{1}{\sqrt{2}} , \nonumber \\
\sqrt{(1-C_{ij})^2 + C_{ij}^2}-(1-C_{ij})\leq N_{ij}\leq C_{ij} \quad {\rm for} \quad C_{ij}>\frac{1}{\sqrt{2}} .
\label{eq_NandC}
\end{eqnarray}
\end{linenomath}
\vspace{-12pt}

From Figure~\ref{fig_3} and Equation~(\ref{eq_NandC}), we can see that all states with a concurrence greater than $1/\sqrt{2}$ violate the Bell-CSHS inequality. We note that this agrees with the Verstraete and Wolf results showing the relation between the nonlocality parameter and the concurrence \cite{VW02b}. For the other values of the concurrence, only some states violate the Bell-CSHS inequality, and the range of negativity guaranteeing the violation of the Bell-CSHS inequality decreases with the decreasing values of the concurrence.
The probability of having a state that violates the Bell-CHSH inequality drops with the decreasing values of the concurrence.

It has been demonstrated that all two-qubit states for which concurrence $C_{ij}>1/\sqrt{2}$ violate the CHSH inequality. This is closely related to the definition of nonlocality parameters, as defined in Equations~(\ref{nonlocality_paramIa}) and (\ref{nonlocality_paramIb}). We can see from these formulas that the nonlocality parameter reaches one of two possible values. For instance, for $M_{12}$, the possible values are $\left(P_{001}-P_{010}-P_{100} \right)^2 + 4P_{010}P_{100}$ and $8P_{010}P_{100}$. It is important to note that, for $C_{12} \leq 1/ \sqrt{2}$, the parameter $M_{12}$ can obtain values exceeding 1 provided that it is determined by $M_{12}=\left(P_{001}-P_{010}-P_{100} \right)^2 + 4P_{010}P_{100}$. Conversely, when $C_{ij}>1/\sqrt{2}$, the expression $8P_{010}P_{100}$ is always greater than 1. However, the nonlocality parameter $M_{12}=8P_{010}P_{100}$ provided that $N_{12}\leq\frac{1}{2}\left( -1+\sqrt{1-C_{ij}^2}+\sqrt{2+3C_{ij}^2-2\sqrt{1-C_{ij}^2}} \right)$. For the case of other values of $N_{12}$, the nonlocality parameter $M_{12}$ is equal to $ \left(P_{001}-P_{010}-P_{100} \right)^2 + 4P_{010}P_{100}$. We observe similar behavior for other nonlocality parameters ($M_{23}$ and $M_{13}$).

\section{Linear Entropy and Degree of Nonlocality}
\label{sec_entropy}
We already know that, for the cases analyzed above, there exist two-qubit mixed states that violate the Bell-CHSH inequality. This suggests that the negativity, concurrence, and the measure of nonlocality should be compared with the degree of mixedness of the mixed two-qubit states. We quantify the mixedness of a state by the linear entropy $ E(\rho) $ determined via its~purity:
\begin{linenomath}
\begin{equation}
E(\rho)\equiv\frac{D}{D-1}\left[1-Tr\left( \rho^2\right) \right] ,
\end{equation}
\end{linenomath}
where $D$ stands for the dimension of $\rho$. For the qubit--qubit states described by the reduced density matrix $\rho_{ij}$ with $D=4$, the linear entropy takes the following form:
\begin{linenomath}
\begin{equation}
E_{ij}\equiv\frac{4}{3}\left[1-Tr\left( \rho_{ij}^2\right) \right] .
\end{equation}
\end{linenomath}

The linear entropy admits values from zero (pure states) to one (maximally mixed~states).

\begin{figure}[H]
\includegraphics[width=0.6\textwidth]{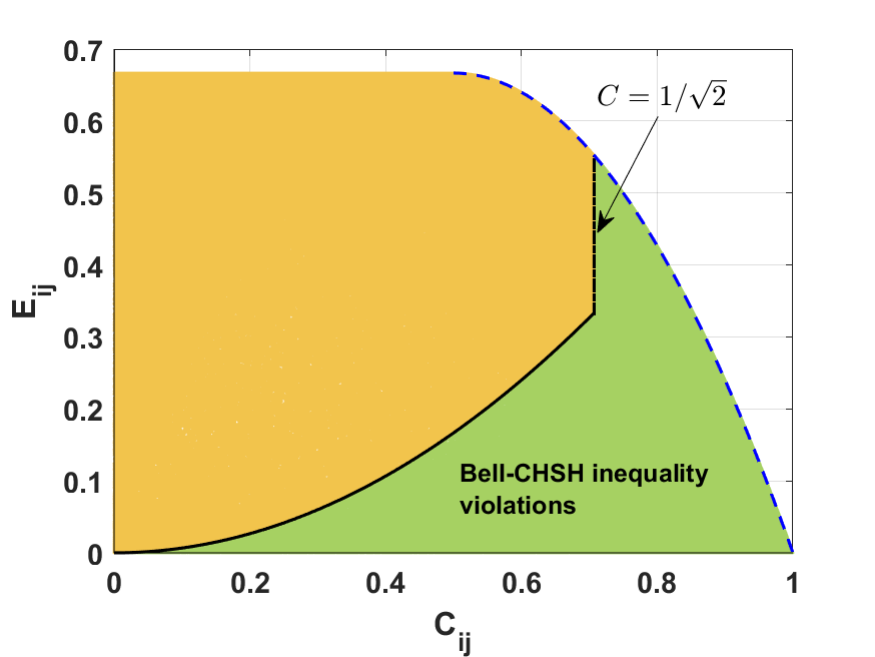}
\caption{Linear entropy $E_{ij}$ versus concurrence $C_{ij}$ for the states violating (green area) and fulfilling (yellow area) the Bell-CSHS inequality established numerically. Black curves (solid and dot-dashed) are plotted according to the derived formulas. The blue dashed curve is plotted using the results of \cite{KL17}.}
\label{fig_4}
\end{figure}

In Figure~\ref{fig_4}, we plot the allowed interval of values of the linear entropy keeping the value of the concurrence fixed for the states violating (green area) and fulfilling (yellow area) the Bell-CSHS inequality. The black solid and dot-dashed lines represent the border between these groups of states. The blue dashed line expresses the maximal value of mixedness for a given value of the concurrence. For $C_{ij} \geq \frac{1}{2}$, the maximal value of linear entropy $E_{ij}$ is described by the relation $8\left(C_{ij}-C_{ij}^2 \right)/3$ (see \cite{KL17}). Moreover, for suitable values of linear entropy and concurrence, we observe only the states that violate the Bell-CSHS inequality. Similarly, as in Figure~\ref{fig_3}, the presented results are obtained~numerically.

To find the relation that describes the border between the discussed groups of states in terms of the concurrence, we derive the relations describing the linear entropy for all pairs of qubits by using the following probabilities (for more details, see \cite{KL17}):
\begin{linenomath}
\begin{eqnarray}
E_{12} \equiv \frac{8}{3}\left(-P_{100}^2+P_{100} -P_{010}^2+P_{010}-2P_{100}P_{010}\right) , \nonumber \\
E_{23} \equiv \frac{8}{3}\left(-P_{001}^2+P_{001} -P_{010}^2+P_{010}-2P_{001}P_{010}\right) , \nonumber \\
E_{13} \equiv \frac{8}{3}\left(-P_{100}^2+P_{100} -P_{001}^2+P_{001}-2P_{100}P_{001}\right) .
\label{eq_E}
\end{eqnarray}
\end{linenomath}

Then, applying Equations~(\ref{nonlocality_paramIa}), (\ref{nonlocality_paramIb}), (\ref{C_ij_probab}), and (\ref{eq_E}), we find the relation that describes the interval of values of the linear entropy keeping the concurrence fixed for the states that violate the Bell-CSHS inequality. This means that, when the nonlocality parameter $M_{ij}>1$, the values of the linear entropy satisfy the following relations:
\begin{linenomath}
\begin{eqnarray}
 E_{ij} < \frac{2}{3}C_{ij}^2 \qquad {\rm for} \qquad C_{ij}\leq\frac{1}{\sqrt{2}} , \nonumber \\
E_{ij} \leq \frac{8}{3}\left(C_{ij}-C_{ij}^2 \right) \qquad {\rm for}
\qquad C_{ij} >\frac{1}{\sqrt{2}}.\label{eq_EandC}
\end{eqnarray}
\end{linenomath}

In Figure~\ref{fig_4}, the black solid line represents the states for which $E_{ij} = 2C_{ij}^2 /3$. Importantly, these are the states obeying $M_{ij}=1$. The blue dashed line corresponds to the states with the negativity taking the minimal value $N_{\rm min}=\sqrt{(1-C_{ij})^2 + C_{ij}^2}-(1-C_{ij})$. Both in the previous section and here, all the states for which the concurrence obeys the relation $C_{ij} > 1/\sqrt{2}$ violate the Bell-CSHS inequality. Also, more and more states violate the inequality when the value of the concurrence increases. Additionally, only the maximally mixed states with concurrence $C_{ij} > 1/\sqrt{2}$ violate the Bell-CSHS inequality.

\begin{figure}[H]
\centering
\hspace{-15em}\includegraphics[width=0.6\textwidth]{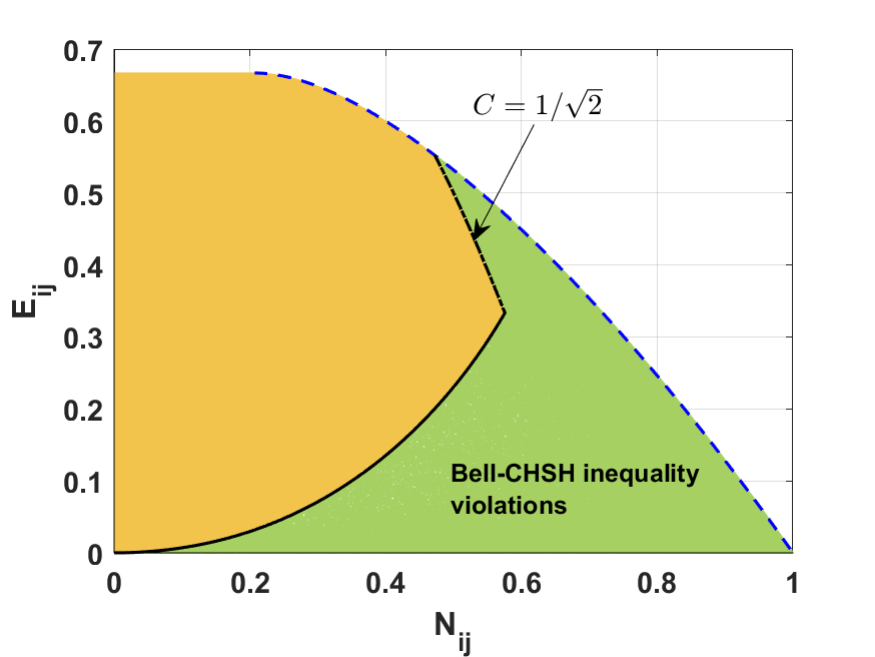}
\caption{Linear entropy $E_{ij}$ versus negativity $N_{ij}$ for the states violating (green area) and fulfilling (yellow area) the Bell-CSHS inequality established numerically. Black curves (solid and dot-dashed) are plotted according to the derived formulas. The blue dashed curve is plotted using the results of \cite{KL17}.}
\label{fig_5}
\end{figure}

Figure~\ref{fig_5} shows the linear entropy for a given value of negativity considering the states that violate and fulfill the Bell-CHSH inequality. The blue dashed curve represents the states with the maximal mixedness parameter value for a given value of negativity. The appropriate formula was derived in \cite{KL17}. The black solid curve belongs to the states where $M_{ij}=1$. For them, applying Equations~(\ref{N_ij_probab}) and (\ref{eq_E}), giving the measure of mixedness and negativity, respectively, we obtain the boundary value of linear entropy for the states which fulfill the Bell-CHSH inequality. We see here that the increase in linear entropy changes the reachable maximal values of the nonlocality parameter $M_{ij}$, and the Bell-CSHS inequality is more strongly violated for the smaller values of $E_{ij}$. The maximal value reachable by the parameter $M_{ij}$ as a function of the linear entropy is determined by the following formula:
\begin{linenomath}
\begin{equation}
E_{ij}=\frac{1}{3}N_{ij} \left(2+N_{ij}-\sqrt{-(2+N_{ij})(-2+3N_{ij})} \right) .
\label{eq_EandN1}
\end{equation}
\end{linenomath}

The black dot-dashed curve characterizes the states with a concurrence equal to $1/\sqrt{2}$. Using this relation and Equations~(\ref{C_ij_probab}), (\ref{N_ij_probab}), and (\ref{eq_E}), the appropriate curve is revealed:
\begin{linenomath}
\begin{equation}
E_{ij}=\frac{\left(1-2N_{ij}^2 \right)\left(-1+4N_{ij}+2N_{ij}^2 \right)}{6N_{ij}^2}.
\label{eq_EandN2}
\end{equation}
\end{linenomath}

Thus, the states violate the Bell-CHSH inequality provided that the linear entropy fulfills the following relations:
\begin{linenomath}
\begin{adjustwidth}{-\extralength}{0cm}
\begin{eqnarray}
E_{ij} < \frac{N_{ij} \left(2+N_{ij}-\sqrt{-(2+N_{ij})(-2+3N_{ij})} \right)}{3} \; &{\rm for}& \; N_{ij} \leq N_I, \nonumber \\
\left. \begin{array}{r}
  E_{ij} < \frac{N_{ij} \left(2+N_{ij}-\sqrt{-(2+N_{ij})(-2+3N_{ij}) } \right)}{3} \\
 \frac{\left(1-2N_{ij}^2 \right)\left(-1+4N_{ij}+2N_{ij}^2 \right) }{6N_{ij}^2}< E_{ij} \leq \frac{-8 \left(-1-N_{ij}+\sqrt{2\left(N_{ij}+N_{ij}^2 \right) } \right) \left(-N_{ij}+\sqrt{2\left(N_{ij}+N_{ij}^2 \right) } \right)}{3}
 \end{array} \right\rbrace \; &{\rm for}& \; N_{II}\leq N_{ij} \leq N_{I}, \nonumber \\
E_{ij}\leq \frac{-8 \left(-1-N_{ij}+\sqrt{2\left(N_{ij}+N_{ij}^2 \right) }
\right) \left(-N_{ij}+\sqrt{2\left(N_{ij}+N_{ij}^2 \right) } \right)}{3} \;
&{\rm for}& \; N_{ij}\geq N_{II},
\label{eq_EandN}
\end{eqnarray}
\end{adjustwidth}
\end{linenomath}
where the parameters $N_I$ and $N_{II}$ are defined as
\begin{linenomath}
\begin{eqnarray}
N_I = \frac{1}{\sqrt{2}}-1+\sqrt{2-\sqrt{2}}, \nonumber \\
N_{II} =\frac{1}{2}\left(\frac{1}{\sqrt{2}}-1+\sqrt{\frac{7}{2}-\sqrt{2}} \right).
\end{eqnarray}
\end{linenomath}

\begin{figure}[H]
\centering
\hspace{-15em}\includegraphics[width=0.6\textwidth]{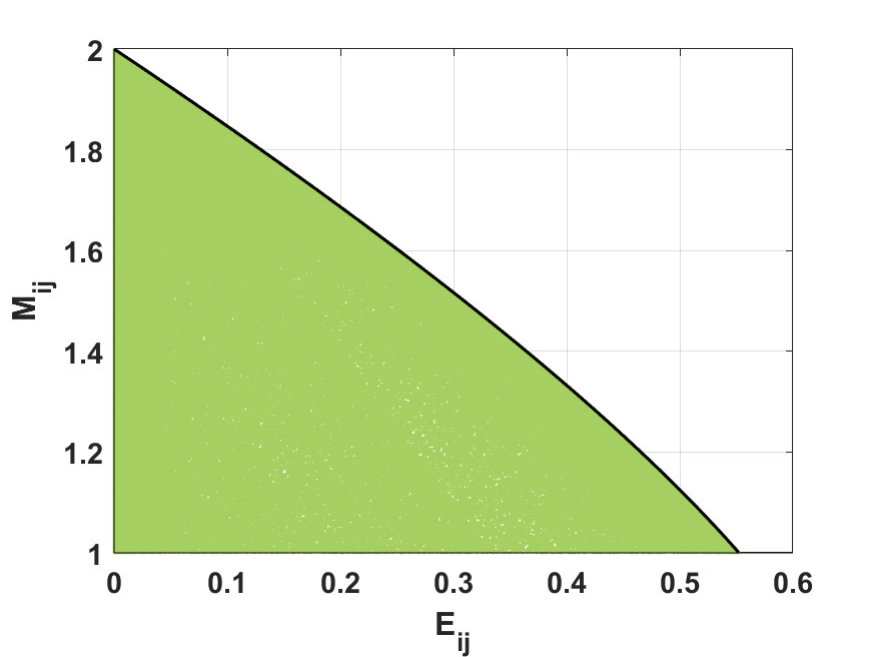}
\caption{Nonlocality parameter $M_{ij}$ versus linear entropy $E_{ij}$ for the states violating the Bell-CSHS inequality established numerically. The black solid curve is plotted according to the derived formulas. }
\label{fig_6}
\end{figure}

Additionally, all states obeying $E_{ij}\geq 4\left(\sqrt{2} - 1 \right)/3 \approx 0.55$ fulfill the Bell-CSHS inequality. The violation of this inequality is observed only for lower values of linear entropy, as documented in Figure~\ref{fig_6}, where the states violating the Bell-CSHS inequality in the plane $(M_{ij}; E_{ij})$ are drawn. We see here that the increase in the linear entropy lowers the reachable maximal values of the nonlocality parameter $M_{ij}$. Also, the Bell-CSHS inequality is more violated for the smaller values of $E_{ij}$. The maximal value of the parameter $M_{ij}$ considered as a function of the linear entropy is determined by the formula
\begin{linenomath}
\begin{equation}
M_{ij}=\frac{1}{8}\left( 2+\sqrt{4-6E_{ij}}\right)^2.
\label{eq_MandE}
\end{equation}
\end{linenomath}

We note that the maximal value of $M_{ij}$ for a given $E_{ij}$ is obtained for the states described by the following two-qubit density matrix:
\begin{linenomath}
\begin{equation}
{\rho_{W}=} \left[ \begin{array}{cccc}
 0 & 0 & 0 & 0 \\
 0 & \alpha/2 & \alpha/2 & 0  \\
 0 & \alpha/2 & \alpha/2 & 0 \\
 0 & 0 & 0 & 1-\alpha
\end{array} \right],
\label{rho_Werner}
\end{equation}
\end{linenomath}
and the corresponding linear entropy is given as
\begin{linenomath}
\begin{equation}
E_{ij}=\frac{8}{3}\left( \alpha - \alpha^2\right).
\label{eq_EforWerner}
\end{equation}
\end{linenomath}
\vspace{-18pt}

\section{Summary}
\label{sec_summary}
The properties of a family of three-qubit states related to quantum nonlocality have been discussed. In particular, we have considered the states with one ($ \langle\hat{n}\rangle=1 $) and two ($\langle\hat{n}\rangle=2$) excitations present in the system, which exhibit the same properties. Using the parameters characterizing nonlocality that were introduced by Horodecki et al. \cite{HHH95,H96}, we have analyzed the conditions that guarantee the violation of the Bell-CHSH inequality by the reduced two-qubit states.

We have analyzed the mutual relation between two-mode nonlocality parameters in three-qubit systems. In particular, we have shown that the two-qubit states both violating and fulfilling the Bell-CHSH inequality can be reached and we have derived the appropriate formulas defining the boundary relations between the two-mode nonlocality parameters. Additionally, it has been proven that, once one of the three two-qubit states violates the Bell-CHSH inequality, the remaining states cannot violate this inequality.

In this paper, the relations among the bipartite negativity, concurrence, parameter of mixedness, and nonlocality parameter have been elucidated. We have found the boundary values of those quantities for the states violating the Bell-CHSH inequality considering two types of states with one and two excitations. We have shown that, provided that the concurrence is greater than $1/\sqrt{2}$, all entangled states violate the Bell-CHSH inequality. We have found the value of negativity parameterized by the concurrence that characterizes the boundary between the regions containing the states violating and fulfilling the Bell-CHSH inequality. Moreover, we have derived similar boundary conditions for the linear entropy drawn as a function of the concurrence and negativity.

\authorcontributions{Conceptualization, J.K.K., W.L., and J.P.J.; methodology, J.K.K.; software, J.K.K.; validation, J.K.K., W.L., and J.P.J.; formal analysis, J.K.K. and W.L.; investigation, J.K.K., W.L., and J.P.J.; writing---original draft preparation, J.K.K. and J.P.J.; writing---review and editing, J.K.K., W.L., and J.P.J. All authors have read and agreed to the published version of the manuscript.}

\funding{J.K.K. and W.L. acknowledge the support provided by the program of the Polish Ministry of Science entitled `Regional Excellence Initiative', project no. RID/SP/0050/2024/1. J.P.~acknowledges the support provided by the project OP JAC CZ.02.01.01/00/22\_008/0004596 of M\v{S}MT.}

\institutionalreview{Not applicable.}

\informedconsent{Not applicable.}

\dataavailability{The raw data supporting the conclusions of this article will be made available by the authors on request.}

\conflictsofinterest{The authors declare no conflicts of interest.}

\begin{adjustwidth}{-\extralength}{0cm}

\reftitle{References}

\end{adjustwidth}
\end{document}